\documentclass[graybox]{svmult}

\usepackage{mathptmx}       		
\usepackage{helvet}         			
\usepackage{courier}        			
\usepackage{type1cm}       	 	
\usepackage{makeidx}         		
\usepackage{graphicx}        		
\usepackage[bottom]{footmisc}		

\makeindex             				

\begin{document}

\title*{PAMOP:  Petascale Atomic, Molecular and Optical Collision Calculations}
\author{B M McLaughlin, C P Ballance, M S Pindzola and A M\"{u}ller}
\institute{B M McLaughlin	\at Centre for Theoretical Atomic, Molecular and Optical Physics (CTAMOP), 
					     School of Mathematics \& Physics, The David Bates Building,
       					     Queen's University, 7 College Park,  Belfast BT7 1NN, UK, \email{b.mclaughlin@qub.ac.uk}
\and C P Ballance \at 	     Department of Physics,  206 Allison Laboratory,
                            			     Auburn University, Auburn, AL 36849, USA \email{ballance@physics.auburn.edu}
\and M S Pindzola \at 	     Department of Physics,  206 Allison Laboratory,
                            			     Auburn University, Auburn, AL 36849, USA \email{pindzola@physics.auburn.edu}
\and A M\"{u}ller  \at 		     Institut f\"{u}r Atom- ~und Molek\"{u}lphysik,
                         			     Justus-Liebig-Universit\"{a}t Giessen, 35392 Giessen, Germany \email{Alfred.Mueller@iamp.physik.uni-giessen.de}
					}
%
%
\maketitle

\abstract*{Petaflop architectures are currently 
		being utilized efficiently to perform large 
		scale computations in Atomic, Molecular 
		and Optical Collisions. We solve the Schr\"odinger or Dirac equation 
		for the appropriate collision problem using the 
		R-matrix or R-matrix with pseudo-states approach.
		We briefly outline the parallel methodology used and 
		implemented for the current suite of Breit-Pauli and DARC codes. In this report, various 
		examples are shown from our theoretical results compared with experimental results 
		obtained from Synchrotron Radiation facilities where the Cray architecture at HLRS
		is playing an integral part in our computational projects.}

\abstract{Petaflop architectures are currently 
		being utilized efficiently to perform large 
		scale computations in Atomic, Molecular 
		and Optical Collisions. We solve the Schr\"odinger or Dirac equation 
		for the appropriate collision problem using the 
		R-matrix or R-matrix with pseudo-states approach.
		We briefly outline the parallel methodology used and 
		implemented for the current suite of Breit-Pauli and DARC codes. In this report, various 
		examples are shown of our theoretical results compared with experimental results 
		obtained from Synchrotron Radiation facilities where the Cray architecture at HLRS
		is playing an integral part in our computational projects.}

\section{Introduction}
\label{subsec:0}
Our research efforts continue to focus on the development of computational 
methods to solve the Schr\"odinger and Dirac equations for atomic and 
molecular collision processes. Access to leadership-class computers such as the 
Cray XE6 at HLRS allows us to benchmark our theoretical solutions against dedicated collision 
experiments at synchrotron facilities such as the Advanced Light Source (ALS), 
Astrid II, BESSY II, SOLEIL and Petra III and to provide atomic and molecular 
data for ongoing research in laboratory and astrophysical plasma science. 
In order to have direct comparisons with experiment, semi-relativistic or 
fully relativistic computations, involving a large number of target-coupled 
states are required to achieve spectroscopic accuracy. These computations 
could not be even attempted without access to HPC resources such 
as those available at leadership computational centers in Europe (HLRS) 
and the USA (NERSC, NICS and ORNL).
We use the R-matrix and R-matrix with pseudo-states (RMPS) methods 
to solve the Schr\"odinger and Dirac equations for atomic and molecular collision processes. 

Satellites such as {\it Chandra} and  {\it XMM-Newton}
are currently providing a wealth of X-ray spectra on many
astronomical objects, but a serious lack
of adequate atomic data, particularly in the {\it K}-shell energy range,
impedes the interpretation of these spectra.
Spectroscopy in the soft X-ray region (0.5--4.5~nm), including
{\it K}-shell transitions of singly and multiply charged ionic
forms of atomic elements such as Be, B, C, N, O, Ne, S and Si, as well as L-shell
transitions of Fe and Ni, provides a valuable probe of the extreme
environments in astrophysical sources such as active galactic nuclei (AGN's),
X-ray binary systems, and cataclysmic variables
\cite{McLaughlin2001,Kallman2010,McLaughlin2013}. For example,
{\it K}-shell photoabsorption cross sections for the carbon isonuclear sequence
have been used to model the Chandra X-ray absorption spectrum of the bright blazar Mkn 421 \cite{McLaughlin2010}.

The motivation for our work is multi-fold; (a) Astrophysical Applications \cite{McLaughlin2010,Phaneuf2011}, 
(b) Fusion and plasma modelling, JET, ITER, (c) Fundamental interest and (d) Support of 
experimental measurements and Satellite observations. 
In the case of heavy atomic systems \cite{Ballance2012,McLaughlin2012},
little atomic data exists and our work  provides results for new frontiers on the application of the 
R-matrix; Breit-Pauli and DARC parallel suite of codes.
Our highly efficient R-matrix codes are widely applicable to
the support of present experiments being performed at  synchrotron radiations facilities 
such as;  ALS, ASTRID II, SOLEIL, PETRA III, BESSY II.
Various examples of large scale calculations are presented to illustrate the predictive nature of the method.

The main question asked of any method is, how do we deal with the many body problem? In our case
we use first principle methods (ab initio) to solve our dynamical equations of motion.  
Ab initio methods provide highly accurate, reliable atomic and 
molecular data (using state-of-the-art techniques) for solving the Schr\"{o}dinger  and Dirac equation.
The R-matrix non-perturbative method is used to model accurately a wide variety of 
atomic, molecular and optical processes such as; electron impact ionization (EII), 
electron impact excitation (EIE), single and double photoionization and inner-shell X-ray processes.
The R-matrix method provides highly  accurate cross sections and rates  used as input 
for astrophysical modeling codes such as; CLOUDY, CHIANTI, AtomDB, XSTAR 
necessary for interpreting experiment/satellite observations of astrophysical objects and 
fusion and plasma modeling for JET and ITER.  

\section{Parallel R-matrix Photoionization}
\label{subsec:1}
The use of massively parallel architectures allows one to attempt calculations which previously could not have been addressed.
This  approach enables large scale relativistic calculations for trans-iron elements such as ; 
Kr-ions, Xe-ions \cite{Ballance2012} and W-ions \cite{Griffin2006,Griffin2013}.
It allows one to provide atomic data in the absence of experiment and takes advantage of the linear algebra libraries available 
on most architectures.  We fill in our Òsea of ignoranceÓ i.e. provide data on atomic elements where none have previously existed.  
The present approach has the capability to cater for Hamiltonian matrices  in excess of 250 K $\times$ 250 K.
Examples are presented for both valence and inner-shell photoionization for systems 
of prime interest to astrophysics and for complex species necessary for plasma modeling in fusion tokamaks.

The development of the dipole codes benefit from similar modifications and developments made to the existing excitation R-matrix codes.
In this case all the eigenvectors from a pair of dipole allowed symmetries are required for bound-free dipole matrix formation.
Every dipole matrix pair  is carried out concurrently with groups of processors assigned to an individual dipole.
The method is applicable to photoionization, dielectronic-recombination or radiation damped excitation and 
now reduces to the time taken for a single dipole formation.  The method so far implemented on various  parallel architectures 
has the capacity to cater for photoionization calculations involving 500 - 1000 levels.  This dramatically improves 
(a) the residual ion structure, (b) ionization potential, (c) resonance structure and  (d) can deal with in excess of 5,000 close-coupled channels. 

 \section{Scalability}
\label{subsec:2}
As regards the scalability of our R-matrix codes, we find from experience on a variety of 
peta-flop architectures that various modules within this suite of codes scale well, upwards to 100,000 cores.
In practical calculations for cross sections on various systems  it is necessary to perform fine energy resolution of resonance
features ( ~10$^{-8}$ (Ry) $\sim$ 1.36 meV) observed in photoionization cross sections. 
This involves many (6 - 30 million) incident photon energies, vital when comparing with  high Ð precision measurements, like 
those for Xe$^{+}$ ions made at the Advanced Light Source synchrotron radiation facility in Berkeley, California, USA 
where energy resolutions of 4 meV FWHM are achieved \cite{Ballance2012}.

The formation of many real symmetric matrices (Hamiltonians),
typically 60 K -150 K, requires anywhere from 10-500 Gb of storage.
The diagonalization of each matrix, from which {\it every} eigenvalue
and {\it every}  eigenvector is required is achieved through use of the ScaLapack package.  
In particular routines : {\bf pdsyevx} and {\bf pdsyevd}, where preference is given to the latter, 
as it ensures orthogonality between all eigenvectors. In typical collision calculations,  
matrices vary in size from 2K $\times$ 2K to 200K $\times$ 200 K, depending
on the complexity of the atomic target.
The formation of the continuum-continuum part of the N+1 electron Hamiltonian is the most time consuming. Therefore 
if there are several thousand scattering channels ({\it nchan}) then there are [{\it nchan} $\times$({\it nchan} +1)/2] matrix blocks. 
Each block represents a partial wave and each subgroup reads a single Hamiltonian and diagonalizes it in parallel,
concurrently with each other. So there is endless scalability.  R-matrix close-coupling calculations are therefore 
reduced to the time required for a single partial wave.  

Optimization of the R-matrix codes on a variety of HPC architectures implements several good coding practices.
We use code inlining (which reduces overhead from calling subroutines), loop unrolling, modularization, unit strides, 
vectorization and dynamic array passing. Highly optimized libraries such as; ScaLapack, BLAS and BLACS 
are used extensively in the codes, which are available at HLRS.  
On Cray architectures, such as the Cray-XE6 at Stuttgart, the R-matrix codes were profiled with performance utilities 
such as  CrayPat which gives the user an indication of their scalability.  
Standard MPI/FORTRAN 90/95 programming is used on the 
world's leading-edge peta-flop high performance supercomputers.
 Serial (legacy) code to parallel R-matrix implementation  
 is carried out  using a small subset of basic MPI commands.

In Table 1 we show details of test runs for the outer region module PSTBF0DAMP for 
K-shell photoionization of B$^{+}$ using 249-coupled states with 400 coupled channels and 409,600 
energy points and an increasing number of CPU cores. A factor of 4 speed up is achieved by using up to 8192 cores.  
The computations were carried out on the Cray-XE6 (Hopper) at NERSC, comparable to the Hermit architecture
at HLRS, Stuttgart using the CrayPat utility.  Note, for actual production runs, 
timings would be a factor of 10 larger, as one would require a mesh of 4,096,000 
energy points to fully resolve the resonances features observed in the spectrum \cite{brendan2014}.
We present the timings for core sizes varying from 1024 to 8192 again for B$^{+}$ K-shell photoionization in its ground state. 
The computations were performed with the outer region module PSTGBF0DAMP for 249-states and 400-coupled channels. 
\begin{table}
\caption{B$^{+}$, 249-states, 400 coupled channels, 409,600 energy points running on an increasing number of cores.
		The results are from the  module PSTGB0FDAMP  for the photoionization cross-section calculations of ground state of the B$^{+}$ ion  
		carried out on the Hopper architectures at NERSC comparable to Hermit at HLRS.  Results  presented illustrate the speed up factor 
		with increasing number of  CPU cores and the total number of core hours.}
\label{tab-photo}       
%
\begin{tabular}{p{2.55cm}p{2.85cm}p{2.85cm}p{2.85cm}}
\hline\noalign{\smallskip}
CRAY-XE6			&Hopper				&Hopper  			&Hopper 		\\
CPU cores			& (NERSC) 			& (NERSC)		& (NERSC)	\\
					& Absolute 			& Speed Up 		& Total		 \\
					& timing (s)			& Factor			& Core Hours	\\
\noalign{\smallskip}\hline\noalign{\smallskip}
1024					&584.19				& 1.0000		 	& 166.1155		 \\
2048					&430.80				& 1.3584			& 245.0077 	\\
4096					&223.08				& 2.6183			&253.8154	      \\
8192					&149.70				& 3.9018			&340.6506	   \\
\noalign{\smallskip}\hline\noalign{\smallskip}
\end{tabular}
\end{table}
The main work horse in our linear algebra code is the ScaLAPACK libraries. 
The goals of the ScaLAPACK project are the same as those of LAPACK; Efficiency (to run as fast as possible), 
Scalability (as the problem size grows so do the numbers of processors grow), Realiability (including error bounds), 
Flexibility (so users can construct new routines from well-designed parts) and
 Ease of Use (by making the interface to LAPACK and ScaLAPACK look as similar as possible).
Many of these goals, particularly portability are  aided by developing and promoting standards, 
 especially for low level communication and computation routines.
 
 Access to the Cray-XE6 at HLRS, January 2013 - May 2013, was provided for exploratory computations, to 
 test our parallel Breit-Pauli and DARC codes.  During this period, Connor Ballance 
 and Brendan McLaughlin, worked closely with Stefan  Andersson, the resident Cray Research consultant at HLRS, 
 profiling and benchmarking our codes on this Cray-XE6.
 A variety of different detailed calculations were made on the Cray-XE6 HLRS 
architecture (Hermit) resulting in several publications \cite{Stolte2013,Soleil2013,Soleil2014}. 
A formal proposal was submitted to HLRS in June 2013 for access to this architecture
 for a project lead by Professor Alfred M\"{u}ller from the University of Giessen. 

\section{X-ray and Inner-Shell Processes}
\subsection{Atomic Oxygen}
\label{subsec:3}
\begin{figure}[t]
\begin{center}
\includegraphics[width=\textwidth]{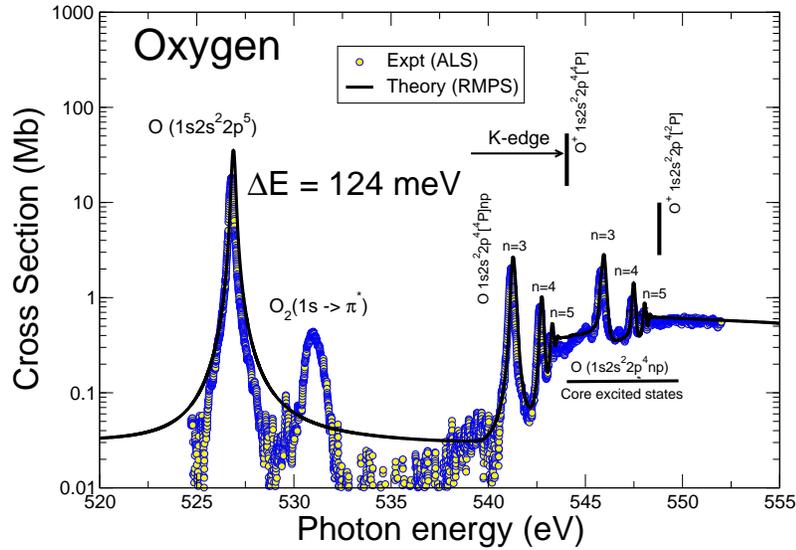}
\caption{\label{fig1} (Colour online) Atomic oxygen photo-absorption cross sections 
								taken at 124 meV FWHM  compared with theoretical estimates. 
								The R-matrix calculations shown are  from the
								R-matrix with pseudo-states method (RMPS: solid black line, present results)
								convoluted with a Gaussian
								 profile of 124 meV FWMH \cite{Stolte2013}.}
\end{center}
\end{figure}

An accurate description of the photoionization/photoabsorption of atomic oxygen is important
for a number of atmospheric and astrophysical applications. 
Photo-absorption of atomic oxygen in the energy region below the $\rm 1s^{-1}$ threshold 
 in X-ray spectroscopy from {\it Chandra} and {\it XMM-Newton} is observed in a 
 variety of X-ray binary spectra.  Photo-absorption cross sections 
 determined from an R-matrix method with pseudo-states (RMPS) and high precision measurements 
 from the Advanced Light Source (ALS) are presented in Fig. \ref{fig1}.
High-resolution spectroscopy with E/$\Delta$E $\approx$ 4,250 $\pm$ 400 were obtained for
photon energies from 520 eV to 555 eV at an energy resolution of 124 $\pm$  12 meV FWHM. 
{\it K}-shell photoabsorption cross-section measurements were made on atomic oxygen at the ALS.
Natural line widths $\Gamma$ are extracted for the $1s^{-1}2s^22p^4 (^4P)np~^3P^{\circ}$  and 
$1s^{-1}2s^22p^4(^2P)np ~^3P^{\circ}$ Rydberg resonances series and compared with theoretical predictions.
Accurate cross sections and line widths are obtained for applications in X-ray astronomy.
Excellent agreement between theory and the ALS measurements is shown which
will have profound implications for the modelling of X-ray spectra and spectral diagnostics. Further details 
can be found in our recent study on this complex \cite{Stolte2013}.

\subsection{Nitrogen ions}
\label{subsec:4}
Recent studies on {\it K}-shell photoionization of neutral nitrogen and oxygen
showed excellent agreement with high resolution measurements made at the Advanced Light Source (ALS)
radiation facility \cite{Stolte2013,Marcelo2011} as have similar cross section calculations on singly and multiply 
ionized stages of atomic nitrogen compared with high resolution 
measurements at the SOLEIL synchrotron facility \cite{Soleil2011,Soleil2013,Soleil2014}. 
 The majority of the high-resolution experimental data from third 
 generation light sources show excellent agreement with 
the state-of-the-art R-matrix method and with other modern theoretical approaches.

The  investigations on Li-like, Be-like and B-like atomic nitrogen ions gives accurate
values of photoionization cross sections produced by X-rays in the vicinity of the {\it K}-edge, 
where strong n=2 inner-shell resonance states are observed. 
N$^{2+}$ ions produced in the  SOLEIL synchrotron radiation experiments are not purely in their ground state (see Figure 2). 
{\it K}-shell photoionization contributes to the ionization balance in a
more complicated way than outer shell photoionization. In
fact {\it K}-shell photoionization when followed by Auger decay couples
three or more ionization stages instead of two in 
the usual equations of ionization equilibrium \cite{Soleil2013,Soleil2014}.

\begin{figure}
\begin{center}
\includegraphics[width=\textwidth]{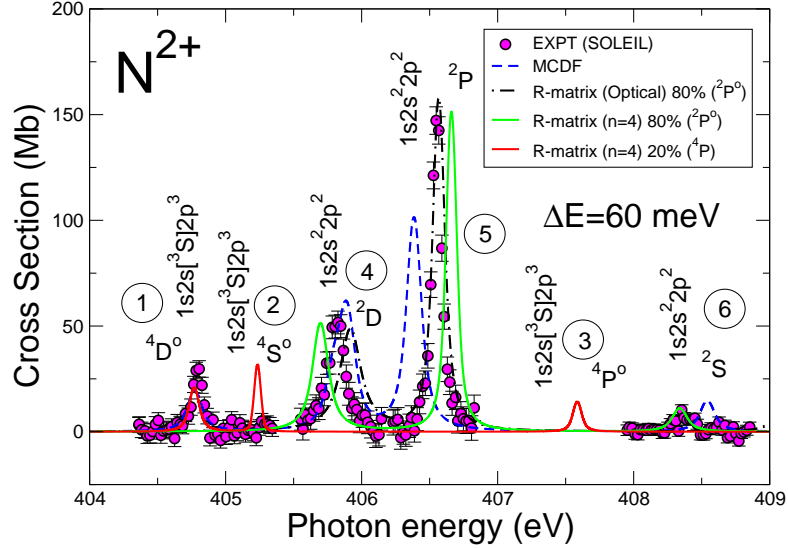}
\caption{\label{fig2}(Colour online) Photoionization cross sections for N$^{2+}$ ions measured 
								with a 60 meV band pass at SOLEIL. 
								Solid circles : total photoionization. 
								The error bars give the statistical uncertainty of the experimental data. 
								The MCDF (dashed line), R-matrix RMPS
								 (solid line, red,  $^4P$, green, $^2P^o$), 
								 and the Optical potential (dash-dot line, $^2P^o$ only) 
								calculations shown were obtained  by convolution with a Gaussian distribution
								 having a profile width at FWHM of 60 meV 
								 and a weighting of the ground and metastable states 
								 to simulate the measurements \cite{Soleil2014}.}  
\end{center}
\end{figure}

 The R-matrix with pseudo-states method (RMPS) was used  to determine all the cross sections (in $LS$ - coupling) with
390 levels of the N$^\mathrm{3+}$ residual ion included in the  close-coupling calculations.   
Since metastable states are present in the parent ion beam,  theoretical PI cross-section calculations are required 
for both the $1s^22s^22p~^2P^o$ ground state and the  $1s^22s2p^2~^4$P metastable 
states of the N$^\mathrm{2+}$ ion for a proper comparison with experiment. 
The scattering wavefunctions were generated by allowing two-electron promotions out of selected base
configurations of N$^\mathrm{2+}$. Scattering calculations were performed with twenty
continuum functions and a boundary radius of 9.4 Bohr radii. For the $^2$P$^o$ ground state and the  $^4P$
metastable states the electron-ion collision problem was solved with a fine energy mesh of 
2$\times$10$^{-7}$ Rydbergs ($\approx$ 2.72 $\mu$eV) to delineate all the resonance features in the PI cross sections.  
Radiation and Auger damping were also included in the cross section calculations.  

\begin{figure}
\begin{center}
\includegraphics[width=\textwidth]{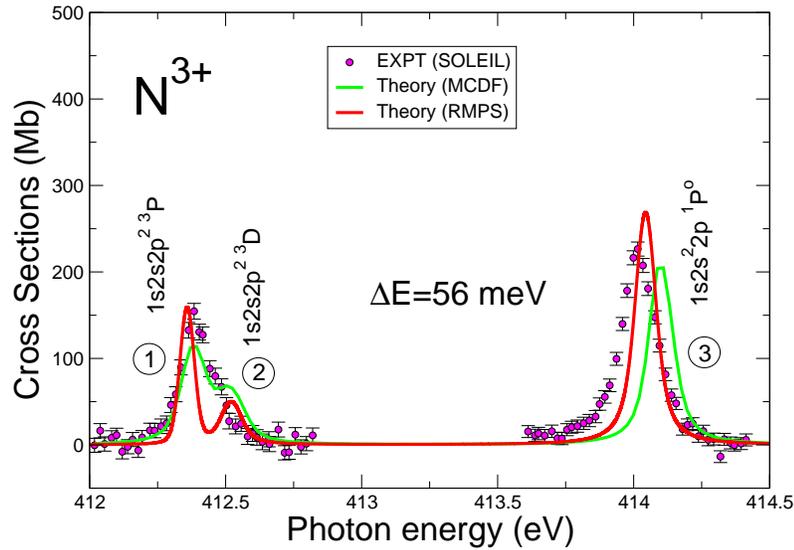}
\caption{\label{fig3}(Colour online) Photoionization cross sections for Be-like atomic nitrogen (N$^{3+}$) ions measured 
								with a 56 meV band pass at the SOLEIL radiation facility. 
								Solid circles :  total photoionization. 
								The error bars give the statistical uncertainty of the experimental data. 
								The MCDF (solid green line) and R-matrix (solid red line) 
								calculations shown are convoluted with a Gaussian
								 profile of 56 meV FWHM and an appropriate weighting of the ground and metastable states 
								 to simulate the measurements. For the metastable $^3P^{\circ}$ state,
								 the MCDF calculations have been shifted up by +1.46 eV in order to match experiment.
								 \cite{Soleil2013}.}
\end{center}
\end{figure}

For a direct comparison with the SOLEIL results,  the R-matrix cross section calculations 
were convoluted with a Gaussian function of appropriate width and 
an admixture of 80\% ground  and 20 \% metastable states used to best  simulate experiment. 
The peaks  found in  the theoretical photoionization cross section 
spectrum were fitted to Fano profiles for overlapping resonances 
as opposed to the energy derivative of the eigenphase sum method \cite{Soleil2013,Soleil2014}.

For Be-like ions both the initial $^1S$ ground state 
and the  $^3$P$^{\circ}$ metastable states were required (see Figure 3).  Cross section calculations were carried 
out in $LS$-coupling with 390-levels  retained in the close-coupling expansion using the R-matrix with 
pseudo states method (RMPS). The Hartree-Fock $1s$, $2s$ and $2p$ 
were used with n=3 physical and n=4 pseudo orbitals of the residual N$^{4+}$ ion.  
 The n=4 pseudo-orbitals were determined by energy optimization on the 
ground state of the N$^{4+}$ ion, with the atomic structure code CIV3.  
The  N$^\mathrm{4+}$ residual 390 ion states used
multi-configuration interaction target wave functions. The non-relativistic
$R$-matrix method determined the energies
of the N${^\mathrm{3+}}$ bound states and all the appropriate cross sections.
We determined  PI cross sections  for the $1s^22s^2$\, $^1$S ground state and the  
 $1s^22s2p$\, $\rm ^3$P$^{\circ}$ metastable  state. 

\begin{figure}
\begin{center}
\includegraphics[width=\textwidth]{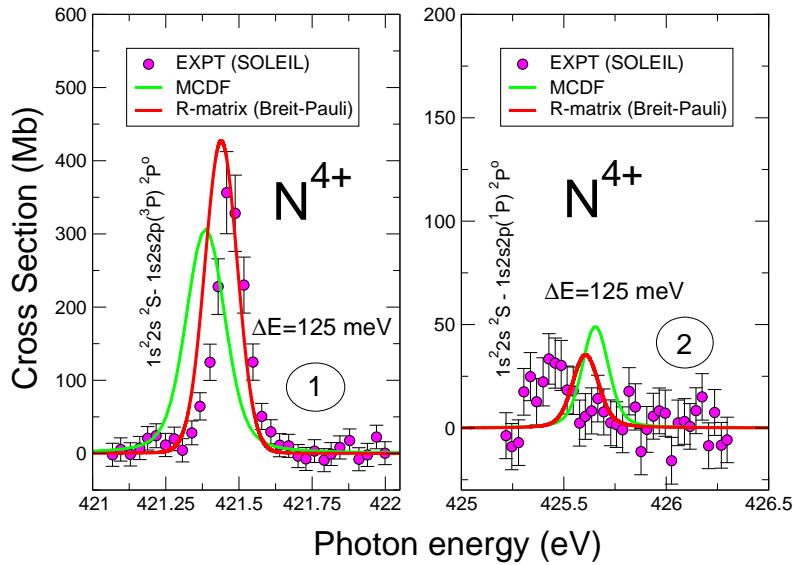}
\caption{\label{fig4}(Colour online) Photoionization cross sections for Li-like atomic nitrogen (N$^{4+}$) ions measured 
								with a 125 meV FWHM band pass at the SOLEIL radiation facility. 
								Solid circles : absolute total photoionization cross sections. 
								The error bars give the  statistical uncertainty of the experimental data. 
								 R-matrix  (solid red line, 31 levels) 
								 intermediate coupling, MCDF (solid green line),
								calculations shown are convolution with a Gaussian
								 profile of 125 meV FWHM to simulate the measurements \cite{Soleil2013}.}
\end{center}
\end{figure}

For Li-like systems, intermediate-coupling photoionization cross section 
calculations were performed using the semi-relativistic Breit-Pauli approximation (see Figure 4). 
An appropriate number of N$\rm ^{5+}$ residual ion states 
(19 $LS$, 31 $LSJ$ levels) were included in our intermediate coupling calculations. 
 The n=4 basis set of  N$\rm ^{5+}$ orbitals obtained  
 from the  atomic-structure code CIV3 were used to represent the wave-functions.  
Photoionization cross-section calculations were then performed in intermediate coupling for the
$1s^22s~^2S_{1/2}$ initial state of the N$\rm ^{4+}$ ion in order to incorporate
relativistic effects via the semi-relativistic Breit-Pauli approximation.

For cross section calculations, He-like $LS$ states were retained:
$1s^2~^1S$,  $1sns~^{1,3}S$, 
$1snp~^{1,3}P^{\,\circ}$, $1snd~^{1,3}D$,  
and $1snf~^{1,3}F^{\,\circ}$, n $\leq$ 4,
of the N$\rm ^{5+}$ ion core giving rise to 31 $LSJ$ states 
in the intermediate  close-coupling expansions for
the $J$=1/2 initial scattering symmetry of the Li-like  N$^{4+}$ ion. 
The n=4 pseudo states are included in an attempt to model core relaxation,  
electron correlations effects and the infinite number of states (bound and continuum) 
left out by the truncation of the close-coupling expansion in our work.
For the structure calculations of the residual
N$\rm ^{5+}$ ion, all n=3 physical orbitals  and n=4 correlation orbitals were included in the 
multi-configuration-interaction target wave-functions expansions used to describe the states. 

\section{Heavy atomic systems}
\subsection{Xe ions}
\label{subsec:5}
Photoionization cross sections of heavy atomic elements, in low stages of ionization,
are currently of interest both experimentally and theoretically and for applications in
astrophysics. The data from such processes have many applications in planetary nebulae,
where they are of use in identifying weak emission lines of $n$-capture elements in NGC
3242. 

Xenon  ions are of importance in man-made plasmas such as XUV light 
sources for semiconductor lithography, ion thrusters for 
space craft propulsion and nuclear fusion plasmas. 
Xenon  ions have also been detected in cosmic objects, 
 e.g., in several planetary nebulae  and in the ejected envelopes 
 of low- and intermediate-mass stars \cite{Ballance2012}.
\begin{figure}
\begin{center}
\includegraphics[width=\textwidth]{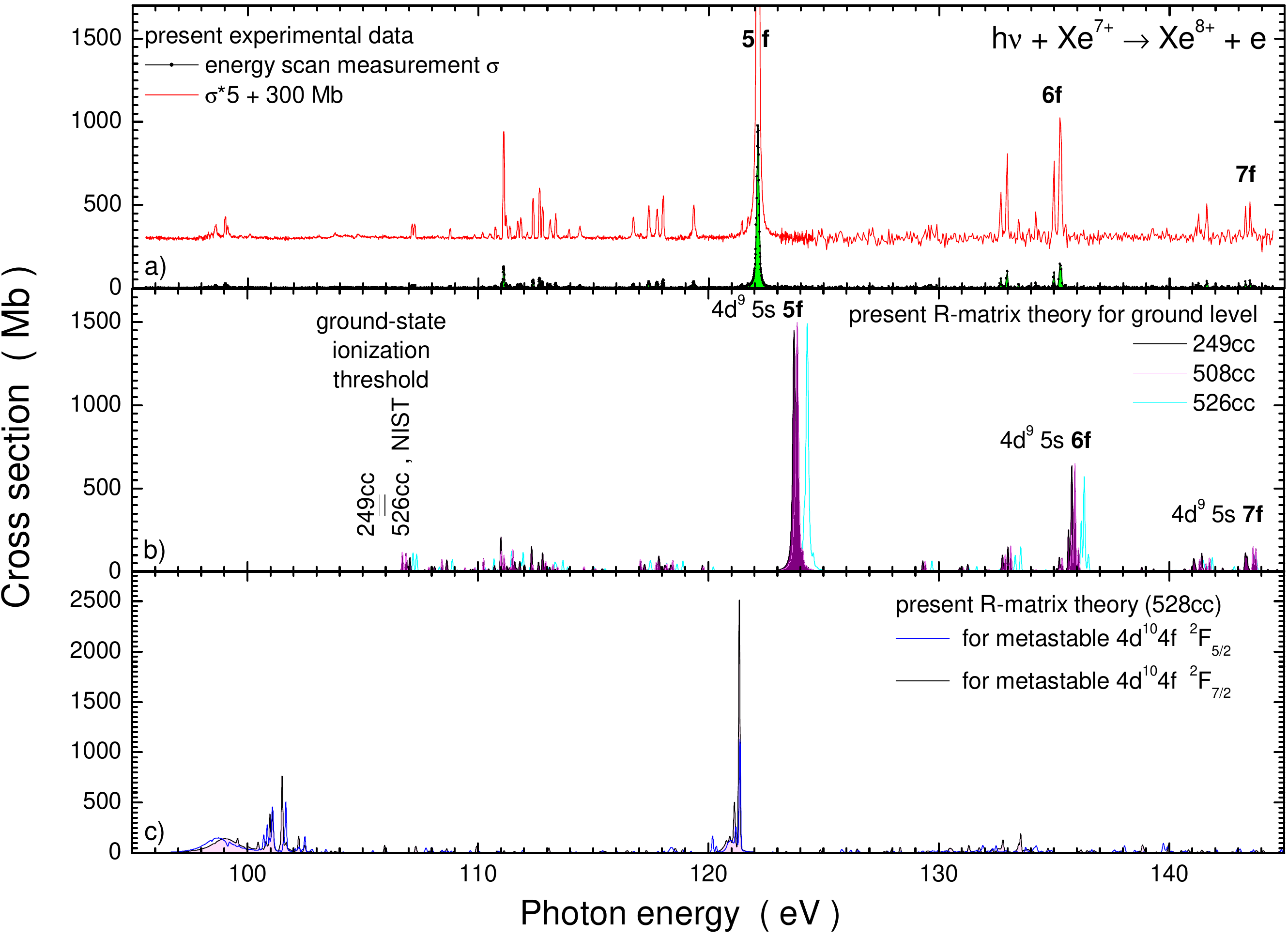}
\caption{\label{fig5}(Colour online)  The experimental and theoretical results
							of the present study of photoionization of Xe$^{7+}$ ions.
							The top (panel a) displays the experimental energy-scan data
							 taken at 65~meV resolution
							(data points connected by straight lines with light shading).
							In order to visualize the contributions of resonances other than
							the dominating $4d^9 5s 5f ~^2{\rm P}^o$ peak, the experimental data
							were multiplied by a factor of 5 and offset by 300~Mb (solid line, red online).
							Panel b) displays the present DARC results for a 249-level, a 508-level and
							a 526-level calculation, abbreviated as 249cc, 508cc and 526cc,
							respectively, for photoionization of ground-state Xe$^{7+}$ ions.
							 Panel c) shows 528-level DARC results for the two metastable
							 fine-structure components of the long lived excited
							 $4d^{10}4f ~^2{\rm F}^o$ term. The theoretical cross sections
							 were convoluted with a 65~meV FWHM Gaussian in order to
							 be comparable with the experiment.}
\end{center}
\end{figure}
Collision processes with highly charged xenon ions are of interest for UV-radiation 
generation in plasma discharges, for fusion research and for space craft propulsion.  
Here we report theoretical and experimental results for the photoionization of Ag-like (Xe$^{7+}$)  xenon ions 
which were measured at the photon-ion end station of ALS on beamline 10.0.1.
Compared with the only previous experimental study of Bizau and co-workers \cite{Bizau2000}  of this reaction, 
the present cross-sections were obtained at higher energy resolution 
 (38 -- 100 meV versus 200 -- 500 meV) and on an absolute cross-section scale. 
 In the experimental photon energy range of 95 -- 145 eV the cross-section 
 is dominated by resonances associated with $4d \rightarrow  5f$ excitation and subsequent autoionization. 
The theoretical results were obtained using the Dirac Coulomb 
R-matrix approximation \cite{venesa2012,Ballance2012,McLaughlin2012}. 
 The small resonances below the ground-state ionization
threshold, located at about 106 eV \cite{Saloman2004}, are due
to the presence of metastable $Xe^{7+} (4d^{10} 4f~^2F^{\circ}_{5/2,7/2})$
ions with an excitation energy of 32.9 eV \cite{Saloman2004} in
the ion beam. In the experimental photon energy range of
95 - 145 eV the cross-section is dominated by resonances
associated with $4d \rightarrow  5f$  excitation
and subsequent autoionization. The most prominent feature in
the measured spectrum is the giant $4d^95s5f ~^2P^{\circ}$ resonance
located at 122.139 $\pm$ 0.01 eV, which reaches a peak cross-section of 1.2 Gb 
  at 38 meV photon energy spread \cite{Schippers2009}.  The experimental
resonance strength of (161 $\pm$ 31)  Mb eV (corresponding
to an absorption oscillator strength of 1.47 $\pm$ 0.28), width of 76 $\pm$3 meV, is
 in suitable agreement with the present theoretical estimate and with previous investigations \cite{Bizau2000}. 
The high-precision cross-section measurements
obtained from the Advanced Light source compared with large-scale 
theoretical calculations obtained from a Dirac Coulomb R-matrix approach 
over the entire photon energy region investigated show suitable agreement.

The present  work on Ag-like ions of Xenon provides a benchmark for future work.
In addition to the direct photoionization process, indirect 
excitation processes  occur for the interaction of a photon with the $4d^{10} 5s~^2S_{1/2}$ 
ground-state  and the metastable $4d^{10} 4f~^2F^{\circ}_{5/2,7/2}$  levels of the Xe$^{7+}$ ion.
These intermediate resonance states can then decay to the ground state or energy accessible excited states.

Photoionization cross-sections on this complex ion were 
performed for the ground ($4d^{10}5s$) and the excited metastable ($4d^{10}4f$)  levels 
associated with the Ag-like xenon ion
to benchmark our theoretical results  with the present high resolution experimental 
measurements made at the Advanced Light Source radiation facility in Berkeley  \cite{Muller2014}.
The atomic structure calculations were carried out using the GRASP code. 
 Initial scattering calculations were performed using 249-levels arising 
  from twelve configurations of the Pd-like (Xe$^{8+}$) residual ion.
  Further collision models were investigated where both a 508-level and 526-level approximation 
  were used in  the close-coupling calculations. For the $4d^{10} 4f~^2F_{7/2,5/2}$ metastable 
  states of this ion, both a 249-level and a 528-level model were investigated \cite{Muller2014}.

 Photoionization cross-section calculations  were performed in 
 the Dirac Coulomb approximation using the DARC codes \cite{venesa2012,Ballance2012,McLaughlin2012} 
for different scattering models for the ground $4d^{10}5s~^2S_{1/2}$ state of the Xe$^{7+}$ ion.  
 The R-matrix boundary radius of 12.03 Bohr radii  was sufficient to envelop
 the radial extent of all the n=6 atomic orbitals of the residual Xe$^{7+}$ ion. A basis of 16 continuum
 orbitals was sufficient to span the incident experimental photon energy
 range from threshold  up to 150 eV. Dipole selection rules for
 the ground-state photoionization require only  the 
 bound-free dipole matrices, $\rm 2J^{\pi}=1^{e} \rightarrow ~2J^{\pi}=1^{\circ},3^{\circ}$. 
 For the excited metastable states,
 $\rm 2J^{\pi}=5^{o} \rightarrow~2J^{\pi}=3^{e}, 5^{e}, 7^{e}$ and
 $\rm 2J^{\pi}=7^{o} \rightarrow~2J^{\pi}=5^{e}, 7^{e}, 9^{e}$ are necessary.

 In the experimental photon energy range of 95 -- 145 eV the cross-section is dominated by resonances
 associated with $4d \rightarrow 5f$ excitation and subsequent autoionization. 
We performed 249-level, 508-level and 526-level DARC photoionization cross-section calculations 
for the Xe$^{7+}$($4d^{10}5s~^2S_{1/2}$) ground state to check on the convergence of our results.   
Figure 5, includes the theoretical cross-sections results  from the 249-state, 508-state,  526-state and 528-state 
models, compared to experiment.  We find best agreement with experiment from an admixture 
of 97.6\% ground and 2.4\% metastable state. 
 Given the complexity of this system, satisfactory agreement with 
 experiment is obtained over the photon energy region investigated.
We see that the small resonances below the ground-state ionization
threshold, occurring at about 106 eV \cite{Saloman2004}, are due
to the presence of metastable $Xe^{7+}(4d^{10} 4f~ ^2F^o_{5/2,7/2})$
ions with an excitation energy of 32.9 eV \cite{Saloman2004} in
the ion beam. 

Recently, the photon-ion merged-beams technique has been employed at the new
Photon-Ion spectrometer at PETRA III (PIPE), in Hamburg, Germany, for measuring
multiple photoionization of Xe$^{q+}$ (q=1 -- 5) ions \cite{Schippers2014}. Prominent ionization features
have been observed in the photon-energy range 650 -- 800 eV, which are associated
with excitation or ionization of an inner-shell 3d electron. Large-scale DARC calculations
are planned for the various charged states of these Xe ions.

\subsection{Tungsten (W) Ions}
\label{subsec:6}
The choice of materials for the plasma facing components in fusion experiments is guided by competing 
desirables: on the one hand the material should have a high thermal conductivity, high threshold for melting 
and sputtering, and low erosion rate under plasma contact, and on the other hand as a plasma impurity it 
should not cause excessive radiative energy loss. The default choice of material for present experiments is 
carbon (or graphite), however tritium is easily trapped in carbon-based walls and for that reason carbon is at present 
held to be unacceptable for use in a D-T fusion experiment such as ITER. 
In its place, tungsten (symbol W, atomic number 74) is the 
favoured material for the wall regions of highest particle and 
heat load in a fusion reactor vessel.

\begin{figure}[t]
\begin{center}
\includegraphics[width=\textwidth]{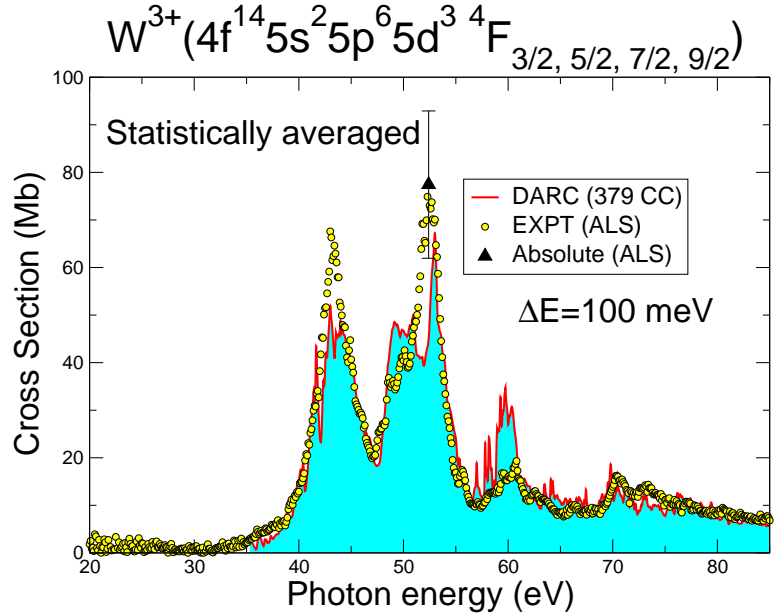}
\caption{\label{fig6} Photoionization of W$^{3+}$ ions over the photon energy range 20 eV - 90 eV. Theoretical work
				(solid red line: DARC) from the 379-level approximation calculations, convoluted with a Gaussian profile of 
				100 meV FWHM and statistically averaged over the fine structure $J$ =3/2, 5/2, 7/2 and 9/2 levels.
				The solid circles (yellow) are from the experimental measurements  made 
				at the ALS using a band width of 100 meV \cite{McLaughlin2014}. The solid triangle is the absolute 
				photoionization cross section accurate to within 20\%. }
\end{center}
\end{figure}

ITER is scheduled to start operation with a W-Be-C wall for a brief initial campaign before switching to W-Be or 
W alone for the main D-D and D-T experimental program. The attractiveness of tungsten
is due to its high thermal conductivity, its high melting point, and its resistance to 
sputtering and erosion, and is in spite of a severe negative factor (that as a high-Z plasma impurity tungsten) does 
not become fully stripped of electrons and radiate copiously, so that the tolerable fraction of tungsten impurity 
in the plasma is at most 2$\times$10$^{-5}$. 

W ions impurities in a fusion plasma causes critical radiation loss and minuscule concentrations prevent ignition.
High resolution experiments are currently available from the ALS on low ionization stages of W ions.
We use the Dirac-Atomic-R-matrix-Codes (DARC) to perform large scale calculations for the single 
photoionization process and compare our results  with experiment.
These systems are an excellent test bed for the photoionization (PI) process where excellent 
agreement is achieved between theory and experiment providing a road-map for electron - impact excitation (EIE).
For photoionization of the W$^{3+}$ ion of tungsten we use a  379 -- level scattering 
model, obtained from 9 configuration state functions of the residual ion.  
Figure \ref{fig6} shows our theoretical results from this 379-level model 
obtained from the DARC codes compared with measurements made at the Advanced Light source \cite{McLaughlin2014}. 
The comparison made in Figure \ref{fig6} illustrates suitable agreement between the theoretical results and the ALS experimental 
measurements which are accurate to within 20\%. 

\begin{acknowledgement}
A M\"{u}ller acknowledges support by 
Deutsche Forschungsgemeinschaft under project number Mu 1068/10  and through
NATO Collaborative Linkage grant 976362 as well as by the US Department of Energy (DOE)
under contract DE-AC03-76SF-00098 and grant  DE-FG02-03ER15424.
M S Pindzola and C P Ballance were supported by US Department of Energy (DOE)
and US National Science Foundation grants  through Auburn University. 
B M McLaughlin acknowledges support from the US National Science Foundation through a grant to ITAMP
at the Harvard-Smithsonian Center for Astrophysics, under the visitor's program, the RTRA network {\it Triangle de le Physique} 
and a visiting research fellowship (VRF) from Queen's University Belfast. 
This research used computational resources at the National Energy Research Scientific
Computing Center in Oakland, CA, USA, the Kraken XT5 facility at the National Institute
for Computational Science (NICS) in Knoxville, TN, USA
and at the High Performance Computing Center Stuttgart (HLRS) of the University of Stuttgart, Stuttgart, Germany.
The Kraken XT5 facility is a resource of the Extreme Science and Engineering Discovery Environment (XSEDE),
which is supported by National Science Foundation grant number OCI-1053575.
The Oak Ridge Leadership Computing Facility 
at the Oak Ridge National Laboratory, provided additional computational resources, 
which is supported by the Office of Science 
of the U.S. Department of Energy under Contract No. DE-AC05-00OR22725.
The Advanced Light Source is supported by the Director,
Office of Science, Office of Basic Energy Sciences,
of the US Department of Energy under Contract No. DE-AC02-05CH11231.
\end{acknowledgement}
%
%
%
%

\end{document}